\documentstyle[11pt,epsfig]{article}

\newcommand{\ome}{\bar{\Omega}_H}

\newcommand{\new}{\newcommand}
\new{\be}{ \begin{equation}}
\new{\ee}{ \end{equation}}
\new{\bea}{ \begin{eqnarray}}
\new{\eea}{ \end{eqnarray}}
\new{\del}{\nabla}
\new{\Del}{\nabla}
\new{\bra}{\left< }
\new{\ket}{\right> }
\new{\E}{{\cal  E}}
\new{\D}{{\cal  D}}
\new{\br}{\bar{r}}
\begin{document}
\begin{flushright}
    KAIST-CHEP-96/2 \\
\end{flushright}

\begin{center}
  \Large{ On the Entropy of a  Quantum Field in 
 the  Rotating  Black Holes}
\end{center}
 
\vspace{2cm}
   
\begin{center}
{Min-Ho Lee\footnote{e-mail~:~mhlee@chep6.kaist.ac.kr
} and Jae Kwan Kim     \\
Department of Physics, Korea Advanced Institute of 
Science and Technology \\
373-1 Kusung-dong, Yusung-ku, Taejon 305-701, Korea.}
\end{center}

\vspace{2cm}

\begin{abstract}
By using the brick wall method we calculate  the  free energy
and the entropy of the  scalar field in the rotating black holes.
As one approaches  the stationary limit surface rather than 
the event horizon in comoving frame, those become divergent. 
Only when  the field is comoving with the black hole (i.e.  
$\Omega_0 = \Omega_H$) those become divergent at the event horizon. 
In the Hartle-Hawking state  the leading terms of the entropy are
$ A \frac{1}{h} + B \ln(h) + finite$, where $h$ is the cut-off in the
radial coordinate near the horizon.
In term of the proper distance cut-off $\epsilon$ it is written as
$ S = N A_H/\epsilon^2$.
The  origin of the divergence is that the density of state
on the stationary surface and beyond it  diverges.   
\end{abstract}
\vspace{1cm}

\newpage
\section{Introduction}
By  comparing  the black hole physics with the thermodynamics  
and discovering of  the black hole evaporation by  Hawking, 
it was shown that the black hole entropy is proportional to the
horizon area\cite{Bekenstein1,Hawking}.
\be
S_{BH} = \frac{A_H}{4}
\ee
in unit $\hbar = c = G = 1$.
In   Euclidean path integration approach  it was shown the tree level 
contribution of the gravitation action gives the black hole  
entropy\cite{Gibbons}.
However the exact statistical origin of the Bekenstein-Hawking 
black hole entropy is unclear.

Recently many efforts have been concentrated  to understand  
the statistical origin of  black hole thermodynamics, specially the 
black hole entropy by various methods (for review 
see \cite{Bekenstein2}): 
't Hooft  was calculated the entropy  of a quantum field propagating
the outside of the black hole.  After the regularization he obtained
 $S = 1/4 A_H$ (the brick wall method)
 \cite{tHooft,Pa1,uglum,barbon}.
Another approach is to identify the black hole entropy 
with the entanglement entropy $S_{ent}$.
Entanglement entropy arises from ignoring the degree of freedom
of a proper region of space: $S = - Tr \rho \ln \rho$.
It is found that the entropy is proportional to the area of the 
boundary\cite{Ent}.
In fact the entanglement entropy and the brick wall method are 
equivalent.
Frolov and Novikov argued that the black hole entropy can be 
obtained  by identifying the dynamical degrees of freedom  
with the states of all fields which are located inside the 
black hole\cite{Frolov}. 
The leading term of the entropy obtained by those methods 
is proportional to the surface area of the  horizon.
However  the proportional coefficient diverges as the cut off goes to 
zero.
The conical approach also gives similar result with 
others\cite{conical}. 
The divergence is because of    an 
infinite number of states near the horizon,
which can be explained by the equivalence principle \cite{Barbon}.
An alternative approach by Frolov is  to identify the black hole 
entropy with the thermodynamic one.  In this approach  the entropy
is finite\cite{Frolov2}.
However they all  treat the only spherical symmetrical black hole.

If the black hole has a rotation, what is changed ?
It is well known that in a  rotating black hole spacetime  a particle
with a zero angular momentum  dropped   from infinity  is dragged 
just by the influence of gravity so that it acquires an angular 
velocity in the same direction in which the black hole rotates.  
The dragging becomes more and more extreme the nearer one approaches
the horizon of the black hole.
This effect is called the dragging of inertial frames\cite{Misner}. 

Thus the  field at equilibrium with the rotating black hole
must also be rotating.  The rotation  is not rigid  but
locally is  different.
So the velocity of the radiation does not exceed the velocity of light. 
However we do not know how to treat the equilibrium state with 
a locally different angular velocity. More precisely  there 
is no global static coordinates.  So we assume that 
the radiation has a rigid rotation $\Omega_0$ small than or equal to
the extremum value of the local rotation. In  a rotating black hole
the extremum value of it is $\Omega_H$, which is the 
angular velocity of the event horizon.

Recently we  considered the black hole entropy by the 
brick wall method in the charged Kerr black hole in\cite{minho} 
and showed the entropy is proportional to the event horizon 
in Hartle-Hawking states.
In this paper to more deep understand the black hole entropy we shall
investigate the black hole entropy  by the brick wall method
in various  stationary black holes: 
the Kaluza-Klein black hole \cite{kaluza} which is the solution of 
the  4-dimensional effective theory reduced from the 5-dimensional 
Kaluza-Klein theory, and the Sen black  hole\cite{sen} which 
is the solution of the Einstein-Maxwell-dilaton-antisymmetric tensor
gauge field theory came from the heteroitic string theory, 
and the Kerr-Newman black hole\cite{kerr} which is the solution of 
the   Einstein-Maxwell theory.

In order to understand the equilibrium state  of the radiation 
(the field) in the rotating black hole  spacetime  in Sec.2  we will 
first consider  the rotating heat bath  in the flat spacetime.
In Sec.3 we will consider the  radiation in equilibrium
state in Rindler spacetime with rotation, which is the most simple
spacetime having the event horizon and a rotation. 
In Sec.4 we will investigate the entropy of the quantum field
in the  stationary black hole background.
We find  the condition to give the finite value  to the free energy
and the entropy. 
In Sec.5  we calculate the entropy in Hartle-Hawking state for  the 
rotating black holes.
Final section is devoted to the summary.

\section{A Rotating Heat Bath }

Let us consider a massless scalar field with a constant angular 
velocity $\Omega_0$ about $z-$axis at thermal equilibrium  
with a temperature  $T = 1/\beta$ in Minkowski 
spacetime, of which line element in cylindrical coordinate is given by
\begin{equation}
ds^2 =  -dt^2 + r^2 d \phi^2 + dr^2 + dz^2.   \label{metric1}
\end{equation}
In this spacetime the positive frequency field mode  can be written as
 $\Phi_{q,m} (x) = f_{q m } (r,z) e^{- i \omega t + i m \phi } $,
 where $q$ denotes a quantum number and $m$ is  the 
 azimuthal quantum number.

For such a equilibrium ensemble of the states of the scalar 
field the partition function is given by
\begin{equation}
Z  =  \sum_{n_q, m} e^{- n_q ( \omega_q - m \Omega_0 ) \beta }    
\end{equation}
and the free energy is given by 
\begin{equation}
 \beta F =   \sum_m \int_0^\infty d \omega  g(\omega, m) 
         \ln  \left( 1 - e^{- \beta (\omega - m \Omega_0)}  \right),
\end{equation}
where $g(\omega ,m)$ is the density of state for 
a fixed $\omega$ and  $m$.

Following 't Hooft we assume that all possible modes of 
a scalar field vanish
at $r = r_1  $ ( $r_1$ is very small.)
and at $r  = L$.  In the WKB approximation with 
$\Phi = e^{i S(r)  - i \omega t + i m \phi + i k z }$  the radial wave  
number $K( x, \omega,m) = \partial_r S $ is given by
\begin{equation}
K^2 ( x, \omega, m) = \omega^2 - \frac{m^2}{r^2} -  k^2.  \label{con1}
\end{equation}
This expression denotes the ellipsoid in momentum phase space  at a 
fixed frequency $\omega$.
The total number of modes with energy less than $\omega$ and a fixed $m$
is obtained by integrating over the volume of phase space, which is 
determined by (\ref{con1})
\begin{eqnarray}
\nonumber
\Gamma (\omega,m) &=& 
     \sum_m \int d \phi dz \int_{r_1}^L dr \frac{1}{\pi} 
     \int dk  K(x,\omega,m)  \\
      &=& \frac{1}{\pi} \sum_m \int d \phi dz \int_{r_1}^L dr  
      \int dk  \left( \omega^2 - \frac{m^2}{r^2} -  k^2 \right)^{1/2}.
      \label{Pvol}
\end{eqnarray}
The integration over $k$ must be carried out over the phase space that
satisfies $ K^2 \ge 0$.  
$\Gamma ( \omega, m)$ can be obtained by investigating
the shape of the expression (\ref{con1}) in momentum phase space.
Thus the free energy, after the integration by  parts, becomes
\begin{eqnarray}
\nonumber
\beta F  &=&  - \beta \sum_m \int_0^\infty d \omega \Gamma ( \omega,m)
         \frac{1}{e^{\beta ( \omega - m \Omega_0 )} - 1 }   \\
 &=&  - \frac{\beta}{2} \int_0^\infty d \omega 
        \int_{- r \omega}^{r \omega} d m
           ( \omega^2 - \frac{m^2}{r^2} ) 
         \frac{1}{e^{\beta ( \omega - m \Omega_0 )} - 1 }, 
\end{eqnarray}
where we  assume that the azimuthal quantum number $m$ is a continuous 
parameter. 
By making the change of variable $ m = r \omega u $ we obtain 
the free energy  
\begin{equation}
\beta F = - \frac{N}{\beta^3} \int d \phi dz \int_{r_1}^L  
\frac{r}{( 1 - v^2)^2} dr,             \label{free1}
\end{equation}
where $N$ is a constant and $v = r \Omega_0$.
Note that as $L$ goes to $1/\Omega_0$ this partition function diverges
as $\gamma^4$, where $\gamma = ( 1 - v^2 )^{-1/2}$.

From the expression (\ref{free1})  it is easy to obtain expressions 
for the energy $E$,  angular momentum $J$, and entropy $S$ of radiation
\begin{eqnarray}
J & = &  \langle m \rangle_{av}  = \frac{1}{\beta} 
    \frac{ \partial}{\partial \Omega_0} (\beta F)  
    = 4 N \frac{1}{\beta^4}  \Omega_0
    \int r^2 \gamma^6  r dr d \phi dz, \\
E & = & \langle \omega \rangle_{av}   = \Omega_0 \cdot J - 
      \frac{\partial}{\partial \beta } ( \beta F) = N \frac{1}{\beta^4} 
       \int ( 3 + v^2 ) \gamma^4 r dr d\phi dz, \\
S & = &  \beta^2 \frac{ \partial}{\partial \beta } F  =  
4 N   \frac{1}{\beta^3} \int \gamma^4  r dr d \phi dz.  \label{entropy1}
\end{eqnarray}
These coincides with those in ref.\cite{zurek}.
Similarly to the free energy  $F$  these expressions $J, E,$ and $S$ 
 diverge as $ L \rightarrow 1/\Omega_0$.
The divergence is related to  the rigid rotation.
In rigid rotating system the velocity of the comoving observer 
grows as one move from the origin to infinity.
So beyond some point the velocity exceeds the velocity of the light.   
This is unphysical.  Thus a  rotating system cannot have the size 
greater than $1/\Omega_0$.
Therefore to obtain a finite value for $J, E,$ and $S$, we must take
$L < 1/\Omega_0$. In such a finite system $\omega > m \Omega_0$.

Now let us consider above problem in co-moving coordinate that are
rotating with angular velocity $\Omega_0$.  The line element in comoving
frame is given by
\begin{equation}
ds^2 = - (1 - \Omega_0^2 r^2 )dt^2 + 2 \Omega_0 r d \phi' dt 
+ dr^2 + dz^2,   \label{metric2}
\end{equation}
where we have used $\phi' = \phi - \Omega_0 t$.
In this coordinate the positive frequency field mode is  written as
$\Phi_{q m} (x) = {\bar f}_{qm} (r,z) e^{ - i \omega' t + i m \phi' }$.

Because in comoving frame the field has no rotation the free energy is 
given by
\begin{equation}
\beta F =  \int_0^\infty d \omega' g' ( \omega') \ln 
       \left( 1 - e^{- \beta \omega'} \right),   \label{free3}
\end{equation}
where $g' (\omega')$ is the density of state for a fixed $\omega'$. 
In WKB approximation the Klein-Gordon equation $\Box \Phi = 0$ yields
the constraint\cite{Mann} 
\begin{equation}
g^{ab} k_a k_b = 0     \label{con2}
\end{equation}
or
\begin{equation}
- ( \omega' - \Omega_0 m)^2 + ( \frac{1}{r^2} m^2 + k^2 + p^2 ) = 0,
\label{con3}
\end{equation}
where $p = \frac{\partial S}{\partial r}$.
In region where $\Omega_0 r < 1$, for a fixed $\omega'$, this expression
represents the ellipsoid in momentum space. Therefore the total number 
of modes with energy less than $\omega'$ is given by
\begin{eqnarray}
\Gamma' ( \omega' ) &=&  \frac{1}{\pi} \sum_m d \phi dz \int dr
\int dk \left(
( \omega' - m \Omega_0)^2 - \frac{m^2}{r^2 } - k^2 \right)^{1/2} 
\label{star} \\
&=& \frac{4}{3} \int d \phi dz \int_{r_1}^L dr 
       \frac{r}{(1 - \Omega_0^2 r^2 )^2} ~\omega^{'3},      \label{pvol}
\end{eqnarray}
which  is the volume of the ellipsoid.
The expression (\ref{star}) is just the same form as Eq.(\ref{Pvol})
when $\omega \rightarrow \omega - m \Omega_0$.
The phase volume (\ref{pvol}) diverges as $L \rightarrow 1/\Omega_0$.
Inserting the expression (\ref{pvol}) into (\ref{free3}) and integrating
we get
\begin{equation}
\beta F = - \frac{N}{\beta^3} \int d \phi dz \int_{r_1}^L  dr
      \frac{r}{( 1- \Omega_0^2 r^2 )^2}.   \label{free4}
\end{equation}
This expression is the same with Eq(\ref{free1}).
From this  we get the energy $E'$ and the entropy $S$:
\begin{eqnarray}
E' &=& \langle \omega' \rangle_{av} =  
      - \frac{\partial}{\partial \beta } (\beta F ) = 3 
    \frac{N}{\beta^4} A \int_{r_1}^L dr 
    \frac{r}{(1 - \Omega_0^2 r^2 )^2},  \\
S &=&  \beta^2 \frac{\partial}{\partial \beta }( \beta F) = 
   4 \frac{N}{\beta^3}  A \int_{r_1}^L dr 
   \frac{r}{(1 - \Omega_0^2 r^2 )^2},
\end{eqnarray}
where $A = \int d \phi dz$.
It is noted that the entropy $S$ is the same with Eq.(\ref{entropy1}) 
and the energy $E'$ is satisfied with  $E' = E - \Omega_0 J$.
This fact show that the coordinate transformation to comoving frame
only change the energy and not change the entropy in WKB approximation.
Thus in the case of calculating the entropy or the free energy 
it is convenient to choose the comoving frame.
It is noted that in co-moving frame the divergence is related to the  
time component $g_{tt}$  of the metric (\ref{metric2}).

\section{A Thermal Bath in Rindler Spacetime with  a Rotation}

In this  section we will consider the thermal equilibrium 
state of the scalar field with the mass $\mu$ and  an uniform 
rotation about $z-$axis in the Rindler spacetime.    
The line element of the Rindler spacetime in cylindrical coordinates 
is  given  by 
\begin{equation}
ds^2 = -  \xi^2  d \eta^2  + d \xi^2 
        + r^2 d \phi^2 + d r^2.    \label{metric3} 
\end{equation}
In this spacetime the event horizon is at $\xi = 0$, and 
$\xi = constant$ represent the trajectory of 
the uniform acceleration\cite{birrell}.
The importance of the Rindler space-time is that in the large
black hole mass limit  the metric of the black space-time reduces
to that of the Rindler space-time\cite{uglum}.

As in Sec.2, the  WKB approximation with 
$\Phi (x) = e^{- i \omega t + i m \phi + i S(\xi, r)}$
yields
\begin{equation}
K^2 ( \xi, r,\omega, m) = 
\frac{\omega^2}{\xi^2} - \frac{1}{r^2} m^2 - p_r^2  - \mu^2,
      \label{con4}
\end{equation}
where $ K = \partial_\xi S$ and $p_r = \partial_r S$.
In this section we will  calculate the free energy by 
using the slightly different method with that in section 2.

It is important  to note that  in WKB approximation the density 
of state $g(\omega,m)$ is determined by the constraint (\ref{con4}),
and that the free energy is singular at $ \omega = m \Omega_0$. 
In particular if $\omega - m \Omega_0 <  0 $ 
the free energy becomes an imaginary number.
However in the WKB  approximation  we can easily see 
$\bar{\omega}  = \omega - m \Omega_0 > 0$ in the region such that
$ \xi -\Omega_0 r > 0$.  But  in the  region such that $ \xi - \Omega_0
r < 0$  it is possible that  $\omega - m \Omega_0  < 0$.
( More details are in Sec.4.)
Therefore to obtain the finite value for  the free energy 
we must require the system to be in the region such 
that $ \xi - \Omega_0 r >0$.  Then the free energy is written as
\begin{eqnarray}
\nonumber
\beta   F &= &  \sum_m \int_{m \Omega_0}^\infty  d \omega g( \omega, m) 
	 \ln \left(   1 - e^{- \beta (\omega - m \Omega_0)} \right)   \\
 \nonumber
	 &=& \int_0^\infty d \omega \sum_m g(\omega + m \Omega_0,m) \ln 
	 \left( 1 - e^{- \beta \omega}  \right)    \\
	 &=& - \beta \int_0^\infty d \omega \frac{1}{e^{\beta \omega } 
	 -1 }
	 \int d m \Gamma ( \omega + m \Omega_0,m),
\end{eqnarray}
where we have integrated by parts and assumed that the quantum number
$m$ is a continuous variable.  
The total number of modes with energy less than $\omega$ 
is obtained by integrating over the volume of phase space
\begin{eqnarray} 
\nonumber
\Gamma(\bar{\omega}) &=& \int d m \Gamma (\omega + m \Omega_0,m)  \\
\nonumber
    &=& \int dm
    \int d \phi dr \int_{r_1}^L d \xi \frac{1}{\pi}
     \int d p_r  K(\xi, r, \omega + m \Omega_0 ,m)  \\
     &=& \frac{1}{\pi} \int dm \int d \phi dr \int_{r_1}^L d \xi
   \int d  p_r \left(
   \frac{\omega^2}{\xi^2}   + \frac{2}{ \xi^2}
    m \Omega_0 \omega  + 
    \frac{m^2 \Omega_0^2}{\xi^2}
   - \frac{1}{r^2} m^2 - p_r^2  - \mu^2
                    \right)^{1/2}.   \label{pvol2}
\end{eqnarray}
The integrations over $m$ and $p_r$ must be carried out over the 
phase space that satisfies $ K^2 ( \omega + m \Omega_0,m) \ge 0$.
After the integration we obtain 
the number of states  with energy less than $\omega$,  
which  is given by 
\begin{equation}
\Gamma (\omega)  = \frac{4}{3} \int d^3 x 
	\frac{\xi r}{\sqrt{( \xi^2 - \Omega_0^2 r^2 )}}
	\left( \frac{\omega^2}{ \xi^2 - \Omega_0^2 r^2 } 
	    - \mu^2 \right)^{3/2}.
\end{equation}
Thus the free energy  becomes
\begin{equation}
\beta F  =  - \frac{4}{3} \beta  \int d^3 x 
\int_{\mu \sqrt{ \xi^2 - \Omega_0^2 r^2 } }^\infty 
d \omega \frac{1}{e^{\beta  \omega } - 1 }   
	\frac{\xi r}{\sqrt{( \xi^2 - \Omega_0^2 r^2 )}}
	\left( \frac{\omega^2 }{ \xi^2 - \Omega_0^2 r^2 } 
	    - \mu^2 \right)^{3/2}.
\end{equation}
For a massless scalar field ( $\mu =0$ )  the free energy becomes
\begin{equation}
\beta F = - \frac{N}{\beta^3} \int d \phi d r \int_{\xi_1}^L d \xi
	\frac{\xi r}{( \xi^2 - \Omega^2 r^2 )^2}.   \label{Free}
\end{equation}
From   this we get the energy $E$, the  angular momentum $J$, 
and the entropy $S$ of  the  field
\bea
J &=& \bra m \ket_{av} =  4 \frac{N}{\beta^4} \Omega_0
\int \frac{r^2}{(  \xi^2 - \Omega_0^2 r^2 )^3} \xi r d \xi dr dz,  \\
E &=& \bra E \ket_{av} = \frac{N}{\beta^4} 
\int \frac{3 \xi^2 + \Omega_0^2 r^2  }{    (  
    \xi^2 - \Omega_0^2 r^2 )^3  } \xi r d \xi dr dz,  \\
S &=& 4 \frac{N}{\beta^3} \int \frac{1}{(  
    \xi^2 - \Omega_0^2 r^2 )^2} \xi r d \xi dr dz.  \label{ent2}
\eea
It is noted that the thermodynamic quantities $F, E$,  and $S$ 
are divergent
as $\xi \rightarrow \Omega_0 r$ rather than the event horizon.
Only in $\Omega_0 = 0$ case the divergence occurs at the horizon
$\xi = 0$. Such a fact can be  easily understand in the co-moving 
frame, of which line element is  given by 
\begin{eqnarray}
ds^2   &=& - \xi^2 d \eta^2 + r^2 ( d \phi' + \Omega_0 d \eta )^2 + 
    d \xi^2 + d r^2 \\
\nonumber
 &=& - ( \xi^2 - \Omega_0^2 r^2 ) d \eta^2 + 2 \Omega_0 r^2 d \eta 
       d \phi' + r^2 d {\phi'}^2 + d \xi^2 + d r^2,     
\end{eqnarray}
where we used $\phi' = \phi - \Omega_0 \eta $.
In this spacetime the event horizon is at $\xi = 0$. In addition to the 
event horizon there is a stationary limit surface 
at $\xi = \Omega_0 r$,
where the Killing vector $\partial_\eta$ becomes  null. That 
surface is the elliptic  hyper-surface\cite{letaw}.  In the 
interval $ 0 < \xi < \Omega_0 r$, the Killing vector is spacelike.
We can also show that the entropy  in the co-moving frame
is the same form  with Eq.(\ref{ent2}).
{\it These facts   imply that the divergence of the thermodynamic 
quantities is deeply  related to the stationary
limit surface in the co-moving frame rather than the event horizon.}


\section{A Entropy of a Scalar Field in a Rotating Black Hole }

\subsection{General Formalism}

Let us consider  a scalar field with mass $\mu$
in  thermal equilibrium at  temperature $1/\beta$  in the 
 rotating black hole background, 
 of which line element is generally given by
\be
ds^2 =  g_{tt}(r, \theta) d t^2  
    + 2 g_{t \phi}( r,\theta) dt d \phi 
    +  g_{\phi \phi}(r, \theta) d\phi^2 
    + g_{rr} (r,\theta) d r^2 
    +  g_{\theta \theta}(r, \theta)   d\theta^2. 
               \label{Metric}  \\
\ee 
This metric has two Killing vector fields: the timelike 
Killing vector $\xi^\mu = (\partial_t)^\mu$ and the axial
Killing vector $\psi^\mu =(\partial_\phi)^\mu$.
The metrics, we concern,  of the Kaluza-Klein, the Sen, and 
the Kerr-Newman black holes are in the appendix.
The properties of those metrics are 
\be
g_{tt} g_{\phi \phi} - g^2_{t \phi} =   - \Delta(r) \sin^2 \theta
\rightarrow 0 
\ee
and
\be
\left( g_{tt} g_{\phi \phi} - g^2_{t \phi}\right) g_{rr}   
\rightarrow  finite
\ee
as one approaches the horizon. 
Another property is that
there are two important surfaces (the event horizon and the
stationary limit surface), and the two surfaces does not coincide.
On the stationary limit surface the Killing vector $\xi^\mu$ vanishes,
and the Killing vector $\xi^\mu + \Omega_H \psi^\mu$ is null  on
the horizon, where $\Omega_H$ is the angular velocity of the
horizon.

The equation of motion of   the  field  with mass $\mu$ and 
arbitrary coupled to the scalar curvature $R(x)$ is
\be
\left[ \Del_\mu   \Del^\mu   -  \xi R - \mu^2 \right] 
\Psi = 0,           \label{equation}
\ee
where  $\xi$ is an arbitrary  constant.
$\xi = 1/6$ and $\mu =0$ case corresponds to the conformally
coupled one.
We assume that the scalar field  is rotating with a constant 
azimuthal angular velocity $\Omega_0$. 
The  associated  conserved quantities are
angular momentum $J$. 
The free energy  of the system is then given by 
\be
   F =   \frac{1}{\beta} \sum_m \int_0^\infty d \E  g(\E, m) \ln 
       \left( 1 - e^{- \beta( \E - m \Omega_0  )}   \right), 
\ee
where $g(\E,m)$ is the density of state for a given $\E$ and $m$.

To evaluate the free energy we will follow 
the brick wall method  of 't Hooft \cite{tHooft}.
Following the brick wall method we impose a small radial cut-off $h$
such that 
\begin{equation}
\Psi  (x) = 0 ~~~~{\rm for  }~~~ r \leq r_H + h,
\end{equation}
where $r_H$  denotes the coordinate of the event horizon.
To remove the infra-red divergence   we also introduce another 
cut-off  $ L \gg r_H$ such that 
\be
\Psi (x) = 0~~~~ {\rm for} ~~~r \geq L.
\ee
It is noted that the brick wall is spherically  symmetric.
In the WKB approximation with $\Psi = 
e^{- i \E t + i m \phi + i S(r, \theta)}$
the  equation (\ref{equation})  yields 
the constraint \cite{Mann}
\begin{equation}
  p_r^2 = \frac{1}{g^{rr}}  \left[
       - g^{tt} \E^2 + 
       2 g^{t \phi} \E  m - g^{ \phi \phi } m^2 
       - g^{ \theta \theta } p_\theta^2 
       - V(x)      \right],         \label{Con1}
\end{equation}
where  $  p_r = \partial_r S$, $ p_\theta = \partial_\theta S$, and
$V(x) = \xi R(x) + \mu^2$.
In WKB approximation it is important to note that
the number of state for a given  $\E$ is determined by  
$p_\theta, p_r$ and $m$.
The number of mode with energy less than $\E$ and  with a fixed
$m$ is obtained by integrating over $p_\theta$ in phase space. 
\bea
\nonumber
\Gamma (\E,m )  &=& \frac{1}{\pi} \int d \phi d \theta \int dr
\int d p_\theta  p_r ( \E, m,x)   \\
&=& \frac{1}{\pi} \int d \phi d \theta \int dr
\int d p_\theta  
  \left[ \frac{1}{g^{rr}}  \left(
       - g^{tt} \E^2 + 
       2 g^{t \phi} \E  m - g^{ \phi \phi } m^2 
       - g^{ \theta \theta } p_\theta^2 
       - V(x)      \right) 
       \right]^{\frac{1}{2}}.        
\eea
The integration over $p_\theta$ must be carried over the phase space
such that $p_r \geq 0$.

In this point we need some remarks.
In a rotating system, in general, there is a  superradiance effect,
which occurs when $ 0 < \E < m \Omega_0$.
For this range of the frequency the free energy $F$ becomes a complex
number. In case $\E = m \Omega_0$ the free energy is divergent.
Therefore  to obtain a real finite value for the free energy $F$,
we must require that $\E > m \Omega_0$. ( For $ 0 < \E < m \Omega_0$
the free energy diverges. See below.)  This requirement say that
we must restrict the system to be in the region 
such that $g_{tt}^{'} \equiv g_{tt} + 2 \Omega_0 g_{t \phi} 
+ \Omega_0^2 g_{\phi \phi} < 0$. 
In this region $ \E - m \Omega_0 >$,
so the free energy is a finite  real  value.
It is easily  showed  as follows.
Let us define $ E = \E - m \Omega_0$.
Then it is written as 
\bea
\nonumber
E &=& \left(
         \frac{g^{t \phi}}{ g^{tt}  } -  \Omega_0 
      \right) m
        + \frac{1}{- g^{tt}} 
     \left[
       \left( 
         g^{t \phi} m 
       \right)^2 
              + \left( - g^{tt}    \right)
       \left(    V + g^{\phi \phi} m^2 + 
          g^{rr} p_r^2 + g^{\theta \theta }p_\theta^2  
       \right)
     \right]^{1/2} \\
  &=& \left( 
         \Omega - \Omega_0 
     \right)  m   + \frac{ -\D }{g_{\phi \phi}}
     \left[
         \frac{1}{-\D} m^2   +   \frac{ g_{\phi \phi} }{- \D }
       \left( V + \frac{p_r^2 }{g_{rr}}
          + \frac{p_\theta^2}{g_{\theta \theta} } 
       \right)
    \right]^{1/2},   \label{Con2}
\eea
where we used 
\be
g^{tt} = \frac{g_{\phi \phi}}{ \D},~~
g^{t \phi} = \frac{ - g_{t \phi}}{\D},~~
g^{\phi \phi} = \frac{ g_{tt}}{\D},
\ee
and  $ \Omega = -\frac{ g_{t \phi}}{g_{\phi \phi}} $.
Here $ - \D = g_{t \phi}^2 - g_{tt}g_{\phi \phi}$.
From Eq.(\ref{Con2}),  for all $m, p_r$ and $p_\theta$, 
one can see the condition such that $ E >0$ is
\be
 \frac{  \sqrt{-\D }  }{ g_{\phi \phi}}  \pm 
\left( \Omega  - \Omega_0    \right) > 0
\ee
or
\be
g_{tt}^{'} \equiv g_{tt} + 2 \Omega_0 g_{t \phi} 
+ \Omega_0^2 g_{\phi \phi} < 0.
\ee
 Therefore in the region such that $ - g_{tt}^{'} >0$ 
 ( called region I)  the free energy is a real, but
 in the region such that $- g_{tt}^{'} < 0$ (called region II)
  the  free energy is complex.
However  in the region I the integration over the momentum 
phase space is convergent. 
But in the region  II the integration over the momentum 
phase is divergent.
 These facts  become more apparent if we investigate  the momentum 
 phase space.
  In the region  I 
the possible points of $p_i$  satisfying $ \E  - \Omega_0 p_\phi  = E$ 
for a given $E$ are located on the following surface
\begin{equation}
\frac{p_r^2}{g_{rr}} + \frac{p_\theta^2}{g_{\theta \theta}} +
      \frac{- {g'}_{tt}}{- \cal D} \left(
              p_\phi + \frac{g_{t \phi }  
           + \Omega_0 g_{\phi \phi}}{{g'}_{tt}} 
           E  \right)^2 
   = \left( \frac{ E^2}{- {g'}_{tt} } 
   - V \right),   \label{ellipsoid}
\end{equation}
which is the ellipsoid,  {\it a compact surface}. 
Here $p_\phi = m$.
So the density of state $g(E)$ for a given $E$ is finite and 
the integrations 
over $p_i$  give a  finite value.
But in the region  II 
the possible points of $p_i$  are located on the following surface
\begin{equation}
\frac{p_r^2}{g_{rr}} + \frac{p_\theta^2}{g_{\theta \theta}} -
        \frac{ {g'}_{tt}}{- \cal D} \left( 
         p_\phi + \frac{g_{t \phi }  + 
         \Omega_0 g_{\phi \phi}}{{g'}_{tt}} 
	 E     \right)^2 
   =  - \left( 
             \frac{ E^2  }{  {g'}_{tt}} 
             + V \right),
\end{equation}
which is  the hyperboloid,   {\it a non-compact surface}. So 
$g(E)$ diverges and the integration over $p_i$  diverges. 
In case of $g^{'}_{tt} = 0$, the possible points   are 
given by the surface
\begin{equation}
\frac{p_r^2}{g_{rr}} + \frac{p_\theta^2}{g_{\theta \theta}} =
\frac{p_\phi - \left( \frac{ g_{\phi \phi} E^2 }{ \cal D }  + V 
            \right)/ \left( \frac{ 2 g_{t \phi} }{ \cal D } E 
	            \right) 
      }{ \frac{- {\cal D} }{ 2 g_{t \phi} E }  
       },
\end{equation}
which is elliptic paraboloid and also $non-compact$. Therefore 
the value of the $p_i$ integration are divergent.
Actually the surface such that ${g'}_{tt} = 0$ is the velocity of the
light surface (VLS). Beyond VLS (in region II) the co-moving 
observer must move  more rapidly  than the
velocity of light. 
Thus we will   assume  that the  system is in the region I.
( For the possible region I  see Sec. 4.2.)
For example, in the case of $\Omega_0 = 0$ the points 
satisfying ${g'}_{tt} =0$ are on the stationary limit 
surface.
The region of the outside (inside) of the stationary 
limit surface corresponds to the region I (II).
In the rotating system in Sec. 2 the region I is $ r < 1/\Omega_0$
and  $ r > 1/\Omega_0$ corresponds to the region II.
In the Rindler spacetime with a rotation
$ \xi > \Omega_0 r$ corresponds to the region I, and $\xi < \Omega_0 r$
to the region II.



With  the assumption that the system is in the region I 
we  can obtain the free energy  as follows
\begin{eqnarray}
\nonumber
  \beta F  &=& \sum_m \int_{m \Omega_0}^\infty d \E  g(\E, m) \ln 
       \left( 1 - e^{- \beta( \E - m \Omega_0 )}   \right)    \\
\nonumber
       &=& \int_0^\infty d\E \sum_m g(\E + m \Omega_0 , m) \ln 
       \left( 1 - e^{- \beta \E }   \right)             \\
       &=&  - \beta \int_0^\infty d\E \frac{1}{e^{\beta \E} - 1}
            \int d m \Gamma (\E + m \Omega_0, m),
\end{eqnarray}
where we have integrated by parts and we assume that the quantum 
number $m$ is a continuous variable.   
The integrations over $m$ and $p_\theta$  yield 
\begin{equation}
F =   - \frac{4 }{3}  
\int d \phi d \theta \int_{r_H + h}^L dr
\int_{V(x) \sqrt{- {g'}_{tt}}}^\infty d\E  
\frac{1}{e^{\beta \E} - 1 }
\frac{ \sqrt{g_4}}{\sqrt{ - {g'}_{tt} }}  \left( 
\frac{\E^2}{ -  {g'}_{tt} }  - V(x)  \right)^{3/2 }.  
\label{freeenergy}
\end{equation} 
In particular when $\Omega_0 = 0$ and non-rotating case 
$g_{t \phi} = 0$, 
the free energy (\ref{freeenergy}) 
coincides with the expression in ref.\cite{tHooft,barbon} and 
it is proportional to the volume of the optical space in the limit
$V(x) = 0$ \cite{optical}.
It is easy to see that the  integrand diverges  as  
$ r_H + h $ or $L$  approach the surface such that $g_{tt}^{'} = 0$.  
In that case  the contribution  of the $V(x)$ can be negligible. 

For a massless and minimally coupled  scalar field case
($\mu =  \xi = 0$) the free energy reduces to 
\begin{equation}
\beta F = - \frac{N}{\beta^3 }  \int d \theta d \phi 
\int_{r_H + h}^L dr \frac{\sqrt{g_4}}{ ( - g^{'}_{tt } )^2  } 
 = - N \int_0^\beta d \tau \int d \theta d \phi 
\int_{r_H + h}^L dr \sqrt{g_4} 
 \frac{1}{ \beta_{local}^4},            \label{freeenergy2}
\end{equation}
where $\beta_{local} = \sqrt{ - g^{'}_{tt } } \beta$ is 
the reciprocal of the local Tolman temperature \cite{Tolman}
in the comoving frame.
This form  is just the free energy of a gas of 
massless particles at local
temperature $1/\beta_{local}$.


From this expression (\ref{freeenergy2})  
it is easy to obtain expressions for 
the total energy $U$,  angular momentum  $J$,  and entropy   $S$ of 
a scalar field
\begin{eqnarray}
J &=&   \langle m \rangle =
       - \frac{1}{\beta} \frac{\partial}{\partial \Omega_0}
( \beta F) =     \frac{4 N}{\beta^4}  \int d \theta d \phi 
\int_{r_H + h}^L dr \frac{\sqrt{g_4}}{ ( - g^{'}_{tt } )^2  }
\frac{ g_{\phi \phi}}{( - {g'}_{tt})} 
\left( \Omega_0 - \Omega \right),  \\
U &=& \langle \E \rangle = \Omega_0 J + \frac{\partial}{\partial \beta} 
( \beta F)   =   \frac{N}{\beta^4}  \int d \theta d \phi 
\int_{r_H + h}^L dr \frac{\sqrt{g_4}}{ ( - g^{'}_{tt } )^2  }
\left[ 
3 + 4 \frac{ \Omega_0 \left( \Omega_0 - \Omega \right)
 g_{\phi \phi } }{( - {g'}_{tt})}  \right],  \\
S &= & \beta^2 \frac{\partial}{\partial \beta } F = 
       \beta ( U - F - \Omega_0 J) = 
       4 \frac{N}{\beta^3 }  \int d \theta d \phi 
        \int_{r_H + h}^L dr \frac{\sqrt{g_4}}{ ( - g^{'}_{tt } )^2  },
\end{eqnarray}
which are also divergent as one approach the surface such 
that $ g_{tt}^{'} =0$.


\subsection{The region such that $ - g_{tt}^{'} > 0$.}

In  this section we study where is the possible region I
for three black hole,  the Kaluza-Klein,
and the Sen, the Kerr-Newman  black holes,
for $ \Omega_0 = \Omega_H, \Omega_0
<  \Omega_H$ and  the extreme case with $\Omega_0 = \Omega_H$. 

\subsubsection{The Kaluza-Klein black hole }
A) $\Omega_0 =\Omega_H$ {\it case}:
In $\Omega_0 = \Omega_H$ case the position of the light 
of velocity surface is exactly found. 
In such a case $g^{'}_{tt}$ can be written as
\bea
g^{'}_{tt} &=& g_{tt} + 2 \Omega_H g_{t\phi} + 
	             \Omega_H^2 g_{\phi \phi}       \\
\nonumber
&=& \frac{\mu^2}{B \Sigma} ( x  - \bar{r}_H ) 
   \left\{
    \frac{ y^2 \sin^2 \theta }{4 \bar{r}_H^2 }
       ( 1 - v^2) x^3 + 
    \frac{ y^2 \sin^2 \theta }{4 \bar{r}_H^2 }   
    \left( 2 - \bar{r}_- (1 - v^2) \right)   x^2  
   \right.   \\
\nonumber
  & &~~ +
   \left[ -1 + \frac{ y^2 \sin^2 \theta }{4 \bar{r}_H^2 } 
	     \left(
   4 + y^2 ( 1- v^2) \cos^2 \theta - 2 \bar{r}_- 
	     \right) 
   \right] x    \\
\nonumber
& & ~~\left.
   + \left[
       \bar{r}_- + \frac{ y^2 \sin^2 \theta }{4 \bar{r}_H^2 } 
	 \left(
      -4 \bar{r}_H - \bar{r}_- y^2 ( 1- v^2)  \cos^2 \theta 
         \right)  
      \right]  
     \right\}             \\
 &\equiv&   
    \frac{\mu^2}{B \Sigma} ( x - \bar{r}_H )  
         \frac{ y^2 \sin^2 \theta }{4 \bar{r}_H^2 } ( 1- v^2)
        \left( 
        x^3 + a_1 x^2  + a_2 x   + a_3 
        \right)
\eea
for $\theta \neq  0$, where $x = \frac{r}{\mu },
y = \frac{a}{\mu}, \bar{r}_H = \frac{r_H}{\mu}$, and 
$ \bar{r}_- = \frac{ r_-}{\mu}$.
From this we can see that there are two  VLS.
One is the horizon ($r = r_H$), and another 
light of velocity surface (call outer VLS) is given by\cite{Table}
\be
r_{VLS} = 2 \mu \sqrt{-Q} \cos \left( 
          \frac{1}{3} \Theta   \right)
 - \frac{1}{3} a_1 \mu,
\ee 
where
\be
\Theta =  {\rm arccos} \left( 
	    \frac{ P}{  \sqrt{ - Q^2}}  \right)
\ee
with 
\be
Q = \frac{ 3 a_2 - a_1^2}{9},~~~
P = \frac{9 a_1 a_2 - 27 a_3 - 2 a_1^3 }{54}.
\ee
In case of the slowly rotating black hole  ($a$ is small)
the VLS  is approximately given by
\be
r_{VLS} \sim 2 \mu \frac{r_H}{a\sqrt{1 - v^2}\sin\theta} - 
\frac{1}{3}
\left( \frac{2 }{1 - v^2} - \frac{r_-}{\mu} \right) \mu,
\ee
which is an open, roughly, cylindrical   surface.
As $v \rightarrow 1$ or $a \rightarrow 0$ the VLS become more 
distant, which came from the fact that as $v \rightarrow 1$ or
$ a \rightarrow 0$ the coordinate angular velocity
$\frac{d \phi}{d t} = - \frac{g_{t \phi}}{g_{\phi \phi}}$ becomes
vanish.
\begin{figure}[bh]
\vspace{0.5cm}
\centerline{
\epsfig{figure=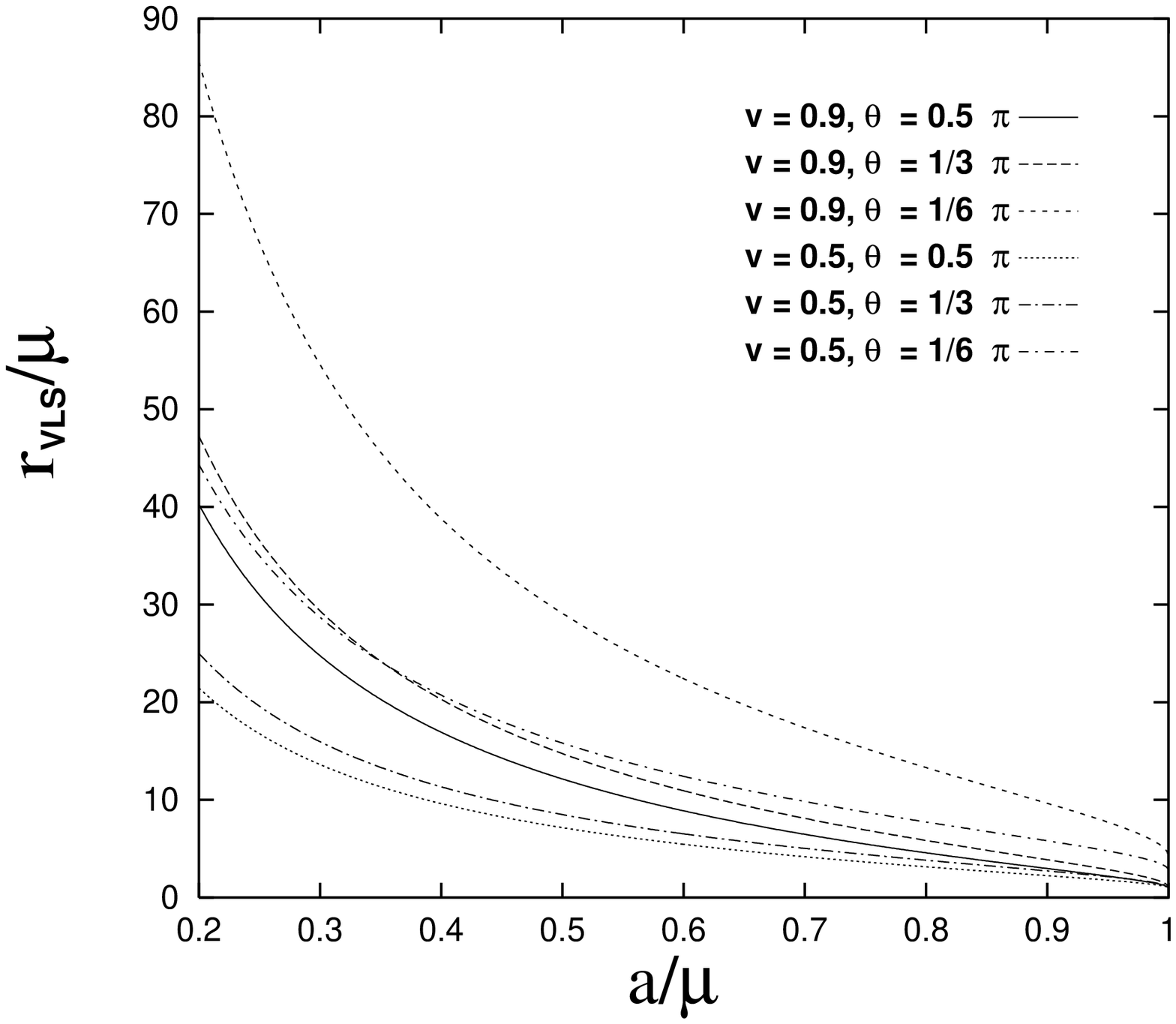,height=8cm, angle=0}}
\vspace{0.1cm}
{\footnotesize Figure 1: The position of the outer velocity of 
light surface for the Kaluza-Klein black hole. }
\end{figure}
For  $\theta = 0$ it is always that $g_{tt}^{'} <0$ for $r >r_H$. 
As $ a \rightarrow \mu$ the outer VLS approaches  horizon. See
Fig.1.

B) $ \Omega_0 < \Omega_H $ {\it case}:
In this case $g^{'}_{tt} = 0 $ is a fourth order polynomial equation 
in $r$ for a given $\theta$.
The region I corresponds to  $ r_{in} < r < r_{VLS}$. 
At $\theta = \pi/2$
$r_{in}$ is  between the stationary limit surface and 
the event horizon, 
and at $\theta = 0$ $r_{in}$ contacts with the event horizon.
 Actually  the inner VLS $r_{in}$  places between the stationary
 limit surface and the event horizon for all $\theta$.
The particular point is that as  $ \Omega_0 \rightarrow \Omega_H$, 
$r_{in}$ approaches  the horizon.
However it does  attach the horizon only when $\Omega_0 = \Omega_H$.
While, the outer velocity of light surface  locates at the very far 
distance from the horizon, and it is a  roughly   cylindrical 
surface as in case $\Omega_0 = \Omega_H$. 
For the  position of the inner VLS  see Fig.2.

C){ {\it  the extreme black hole case with }}  $\Omega_0 = \Omega_H $: 
The  extreme black hole for the Kaluza-Klein black hole occurs when 
$ \mu^2 = a^2$. In this  case  the inner horizon and outer horizon are 
at the same place.
At $\theta = 1/2 \pi$, $g_{tt}^{'}$ is written as
\be
g^{'}_{tt} =  \frac{\mu^2}{B \Sigma} 
( x - \bar{r}_H )^2 x \left( 
       x +   \frac{2}{1 - v^2}    \right)
\frac{ 1 - v^2 }{4},
\ee
which shows  that the possible region such that $g^{'}_{tt} < 0 $ 
does not exist at $\theta = 1/2 \pi$.
Therefore in the extreme black hole case it is impossible to consider
the brick wall model of 't Hooft.
\begin{figure}[bh]
\vspace{0.5cm}
\centerline{
\epsfig{figure=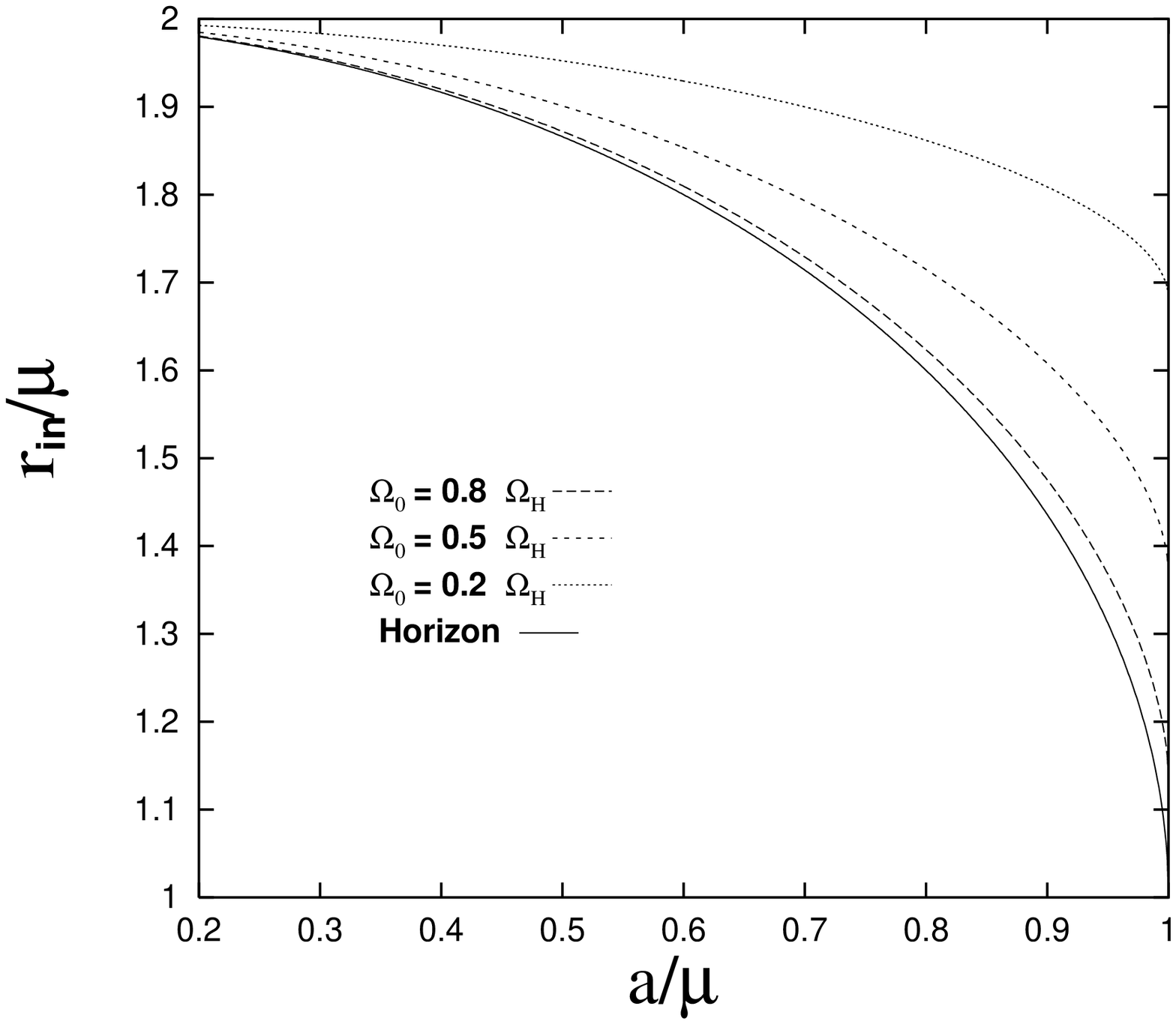, height=8 cm, angle=0}}
\vspace{0.1cm}
Figure 2: The position of $ r_{in}$ at $ \theta = 0.5 \pi$ for the 
Kaluza-Klein black hole. $v = 0.5$.
\end{figure}


\subsubsection{The Sen black  hole}

A) $\Omega_0 = \Omega_H$ {\it case}:
In $\Omega_0 = \Omega_H$ case 
$g^{'}_{tt}$ can be written as
\bea
g^{'}_{tt} &=& g_{tt} + 2 \Omega_H g_{t\phi} + 
	             \Omega_H^2 g_{\phi \phi}       \\
\nonumber
&=& \frac{\mu^2}{ \Sigma} ( x  - \bar{r}_H ) 
   \left\{
    \frac{ y^2 \sin^2 \theta }{4  \bar{r}_H^2  \cosh^4 \gamma}
        x^3 + 
    \frac{ y^2 \sin^2 \theta }{4 \bar{r}_H^2 \cosh^4 \gamma}   
    \left( 2 \cosh 2 \gamma - \bar{r}_-  \right)   x^2  
   \right.   \\
 \nonumber
  & &~~ +
   \left[ -1 + \frac{ y^2 \sin^2 \theta }{ \bar{r}_H^2 } 
    + \frac{ y^2 \sin^2 \theta }{4 \bar{r}_H^2 \cosh^4 \gamma}   
	     \left(
    y^2  \cos^2 \theta - 2 \bar{r}_-  \cosh 2 \gamma
	     \right) 
   \right] x    \\
\nonumber
& & ~~\left.
   + \left[
       \bar{r}_- + \frac{ y^2 \sin^2 \theta }{ \bar{r}_H } 
    - \frac{ y^2 \sin^2 \theta }{4 \bar{r}_H^2 \cosh^4 \gamma}   
	 \left(
       \bar{r}_-  y^2  \cos^2 \theta 
         \right)  
      \right]  
     \right\}             \\
 &\equiv&   
    \frac{\mu^2}{ \Sigma} ( x - \bar{r}_H )  
         \frac{ y^2 \sin^2 \theta }{4 \bar{r}_H^2 \cosh^4 \gamma} 
        \left( 
        x^3 + a_1 x^2  + a_2 x   + a_3 
        \right)
\eea
for $\theta \neq  0$, where $x = \frac{r}{\mu },
y = \frac{a}{\mu}, \bar{r}_H = \frac{r_H}{\mu}$, and 
$ \bar{r}_- = \frac{ r_-}{\mu}$.
Then the exact position  of the inner VLS and outer VLS are
are given by
\be
r_{in} = r_H,~~~ 
r_{VLS} =  2 \mu \sqrt{-Q} \cos \left( 
          \frac{1}{3} \Theta   \right)
 - \frac{1}{3} a_1 \mu.
\ee 
The position of the outer VLS for small $a$ is approximately given by
\be
r_{VLS} \sim \frac{2 \mu r_H \cosh^2 \gamma}{a
 \sin \theta }
 - \frac{ 1}{3} \left( 2 \cosh(2\gamma) - \frac{r_-}{\mu} \right)  \mu,
 \ee
which is an open, roughly, cylindrical surface.
As  $ a \rightarrow 0 $  the VLS goes to  the infinity, and it 
disappears when $a = 0$.
As  $ \gamma$ or $a $ is  increasing   the VLS approaches the horizon.
At $ \theta = \frac{1}{2} \pi$, similarly to the 
Kaluza-Klein black  hole, $g^{'}_{tt} < 0$ for $ r > r_H$.
See Fig.3.
\begin{figure}[bh]
\vspace{0.5cm}
\centerline{
\epsfig{figure= 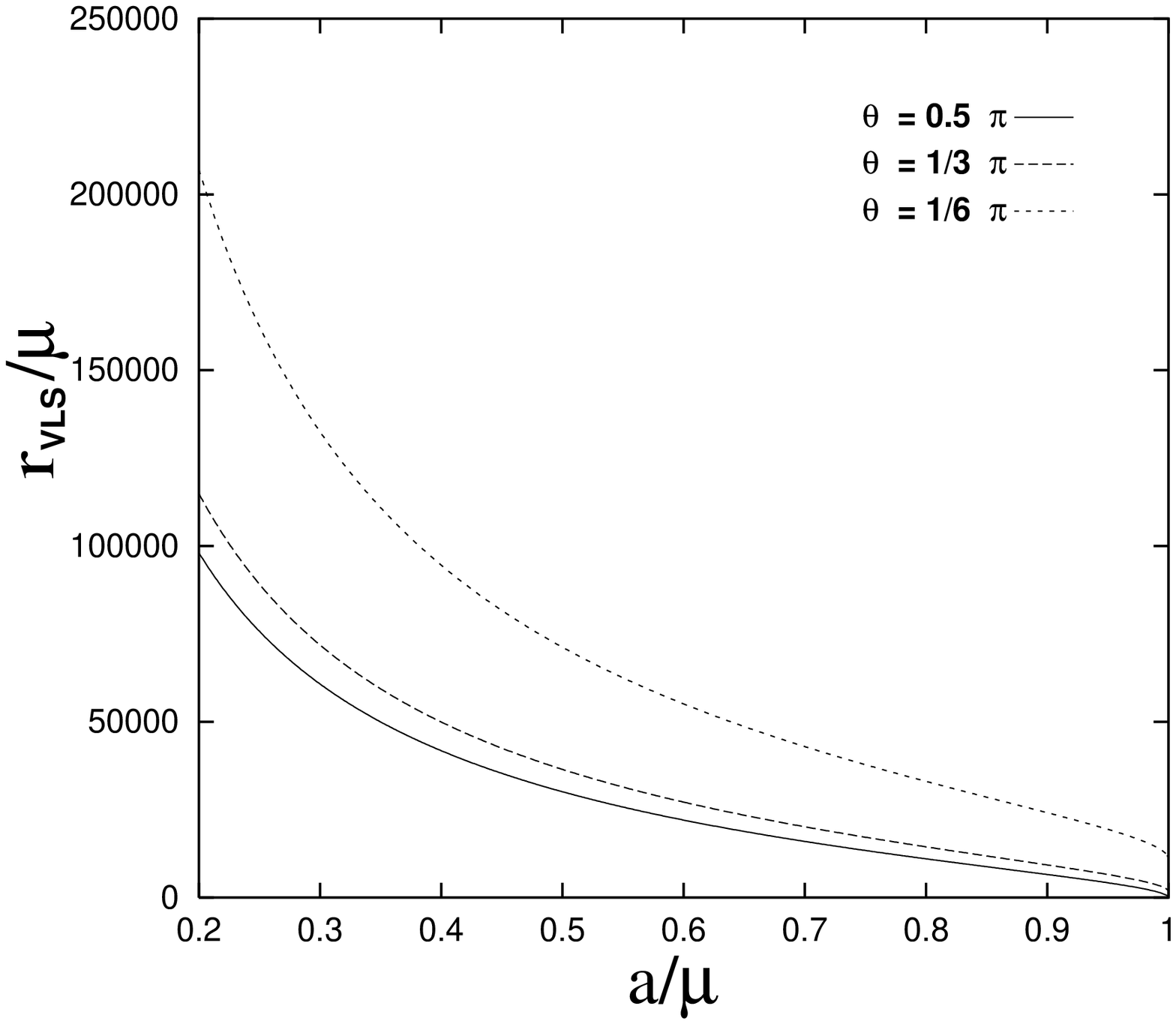 , height=8 cm, angle=0}}
\vspace{0.1cm}
Figure 3: The position of the outer velocity of light surface for the 
Sen black hole. $\gamma = 5.0$.
\end{figure}

B) $ \Omega_0 < \Omega_H $ {\it case}:
In this case $g^{'}_{tt} = 0 $ is also a fourth order  equation 
in $r$ for a given $\theta$.
Similarly to the Kaluza-Klein black hole  the region I is 
$ r_{in} < r < r_{VLS}$. 
At $\theta = 0 $ the inner VLS $r_{in}$ is at the horizon,
and at $\theta = \pi/2$
$r_{in}$ locates  at the between the stationary limit surface and 
the event horizon. See Figure 4. 
As $ \Omega_0 \rightarrow \Omega_H$,  $r_{in} $ approaches 
to the horizon. Only when $\Omega_0 = \Omega_H$ it coincides with the
event horizon.
The outer velocity of light surface, in case of small $a$,  locates
at the very far distance from the horizon, 
and it is a roughly   cylindrical 
surface.
\begin{figure}[bh]
\vspace{0.5cm}
\centerline{
\epsfig{figure=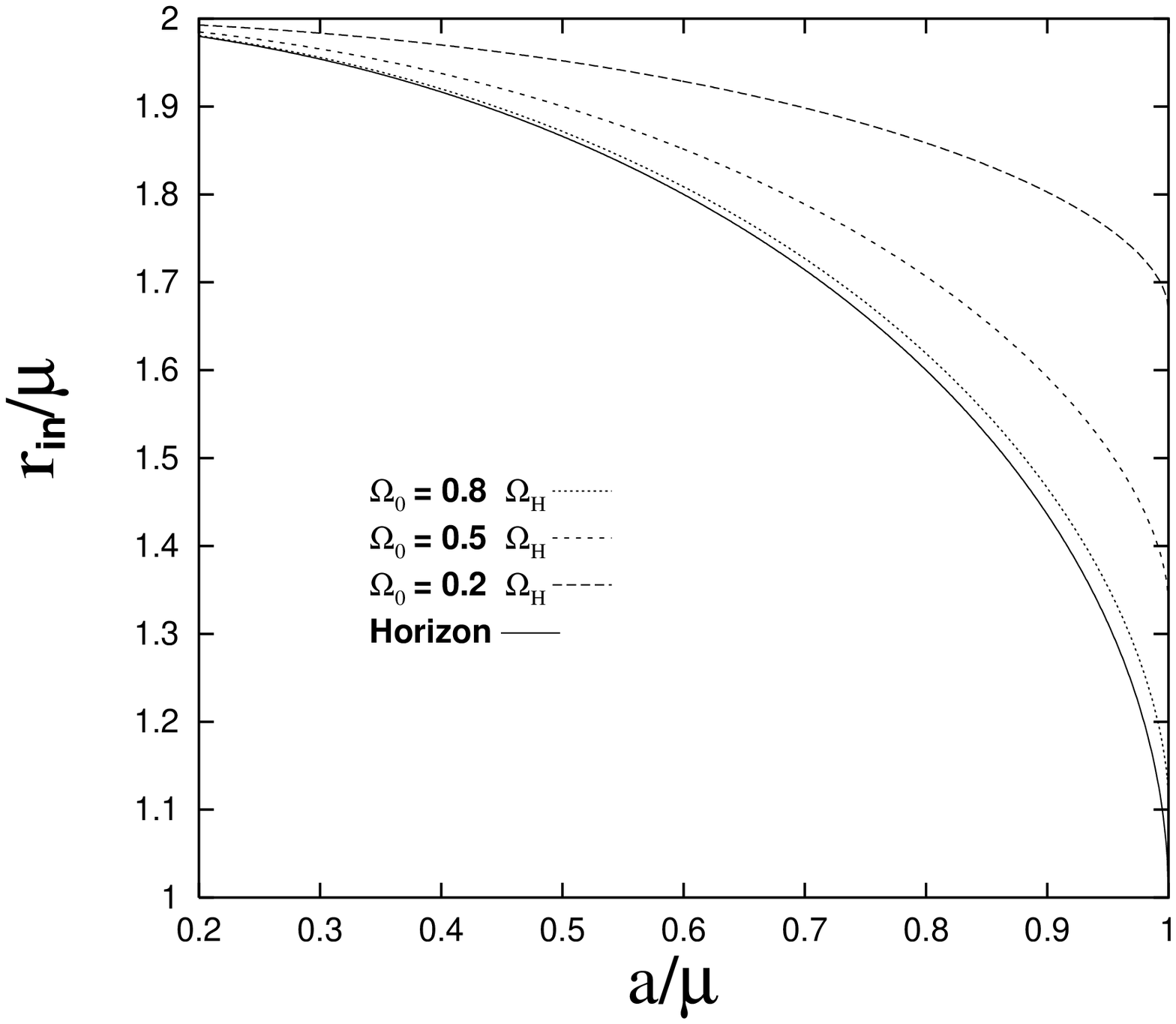, height= 8cm, angle=0}}
\vspace{0.1cm}
Figure 4: The position of the inner velocity of light surface for 
the Sen black hole. $\gamma = 5.0, \theta = 0.5 \pi$. 
\end{figure}

C){ {\it  the extreme black hole case with }}  $\Omega_0 = \Omega_H $: 
The  extreme black hole for the Kaluza-Klein black hole occurs when 
$ \mu^2 = a^2$. 
In this  case  the inner horizon 
and outer horizon are at the same place.
At $\theta = \frac{1}{2} \pi $  $g_{tt}^{'}$  is written as
\be
g^{'}_{tt} =  \frac{\mu^2}{ \Sigma}  
( x - \bar{r}_H )^2 x \left( 
       x +  2 \cosh 2 \gamma  \right)
\ee
which shows that the possible region such that $g^{'}_{tt} < 0 $ 
does not exist at $\theta = 1/2 \pi$.
Therefore in the extreme black hole case it is impossible to consider
the brick wall model of 't Hooft.


\subsubsection{The Kerr-Newman black hole}

A) $\Omega_0 = \Omega_H$ {\it case}:
In  $\Omega_0  = \Omega_H$ case we can exactly find  the 
position of the  light of velocity surface.  
In such a case ${g'}_{tt}$ can be written as
\begin{eqnarray}
{g'}_{tt} &=& g_{tt} + 2 \Omega_H g_{t \phi} + \Omega_H^2 
	         g_{\phi \phi} \\
\nonumber
	  &=& \frac{M^2}{\Sigma}  ( x - \br_H)  \left\{
	  \bar{\Omega}_H^2  \sin^2 \theta ~x^3 + \bar{r}_H
	  \bar{\Omega}_H^2 \sin^2 \theta ~x^2    \right.    \\
\nonumber
	 & & ~ +   \left[ -1 + \ome^2 \sin^2 \theta   \left(
	  y^2 + y^2 \cos^2 \theta  + \bar{r}_H^2    \right)    
	       \right]  x  \\
  &+ & \left.  \left[     
  2 \left( 1 - \ome y \sin^2 \theta \right)^2 - \br_H + 
 \br_H \ome^2 \sin^2 \theta  \left( \br_H^2 + 
 y^2  + y^2 \cos^2 \theta   \right)
 \right] \right\}    \\
	  &\equiv& \frac{M^2}{\Sigma}  ( x - \br_H ) 
	   \ome^2 \sin^2 \theta  \left(
	    x^3  + a_1 x^2  + a_2 x + a_3  \right)
\end{eqnarray}
for $\theta \neq 0$, where
$x = r/M, y = a/M, z = e/M, \ome = M \Omega_H, \br_H = r_H /M $.
Then the exact position of the outer light of velocity surface 
is given by 
\be
 r_{VLS}   =  2 M \sqrt{
 - Q} \cos \left( \frac{1}{3} \Theta  \right) - \frac{1}{3} a_1 M.
 \label{sol}
 \ee
For small $a$  Eq. (\ref{sol})   is approximately  given by 
 \be
 r_{VLS} \sim \frac{1}{\Omega_H \sin \theta } - \frac{r_H}{3  },
 \ee
 which is an open, roughly, cylindrical surface.
 For $\theta = 0$ it is always that  ${g'}_{tt} < 0 $ for 
 $r > r_H$.  
 As $ a \rightarrow 0$, $r_{VLS}$ goes to infinity, and
 as $ a \rightarrow \sqrt{M^2 + e^2 }$ it approaches 
  the event horizon. See Fig.5.
 The inner VLS $r_{in}$ is the event horizon.
\begin{figure}[bh]
\vspace{0.5cm}
\centerline{
\epsfig{figure= 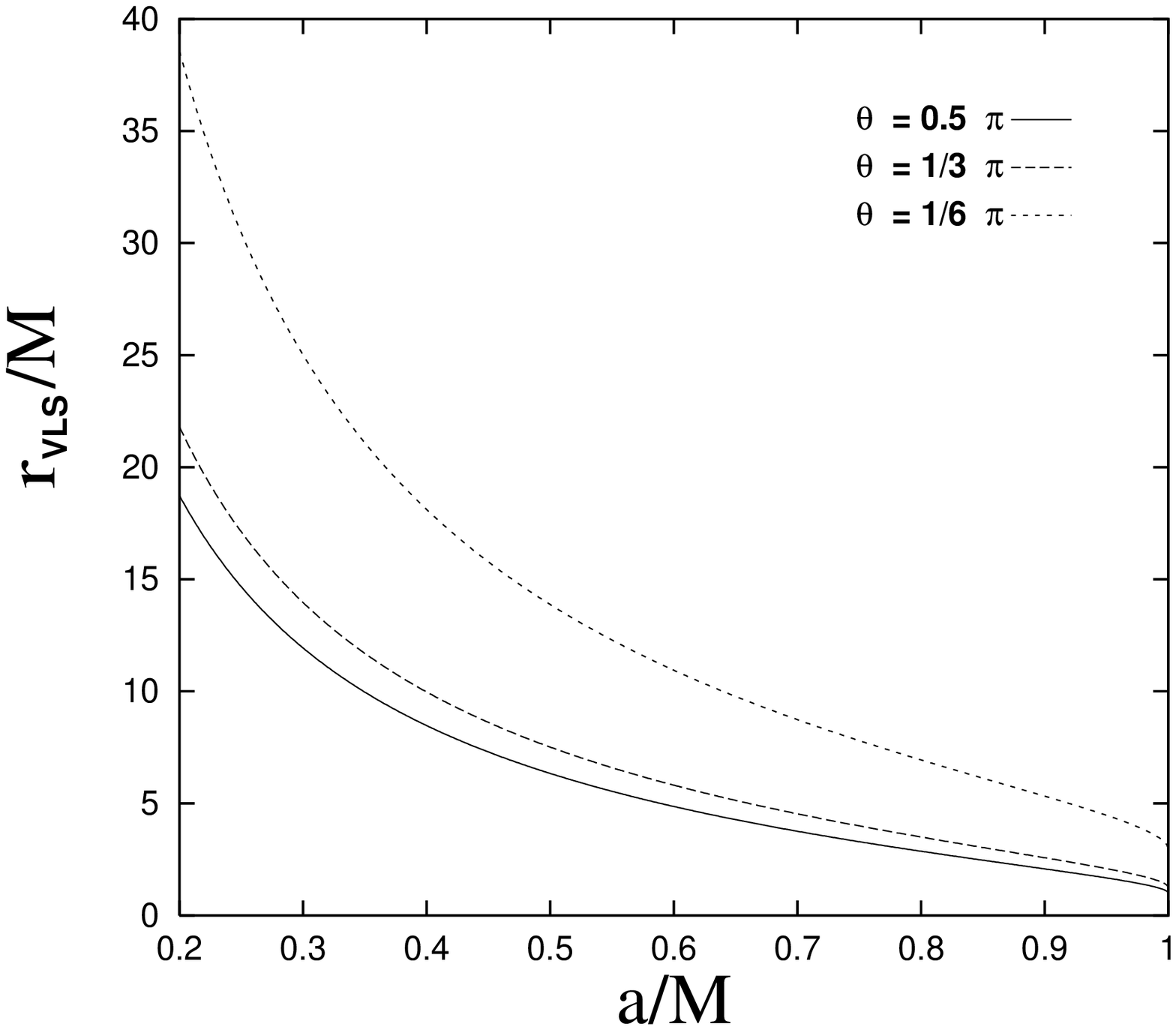, height=8 cm, angle=0}}
\vspace{0.1cm}
Figure 5: The position of the outer  light of velocity surface
for the  Kerr-Newman black hole. $e = 0.0$. 
\end{figure}

B) $\Omega_0 < \Omega_H$ {\it case}:
In  this case, similarly to other black holes, 
the inner VLS $r_{in}$ approaches to the horizon as 
$\Omega_0 \rightarrow \Omega_H$. See Fig.6. 
The inner VLS is a compact  surface, which shrink to horizon as
$\Omega_0 \rightarrow \Omega_J$. See Fig.7.
The outer VLS is at far place, which disappears when $\Omega_0 = 0$.
\begin{figure}[bh]
\vspace{0.5cm}
\centerline{
\epsfig{figure= 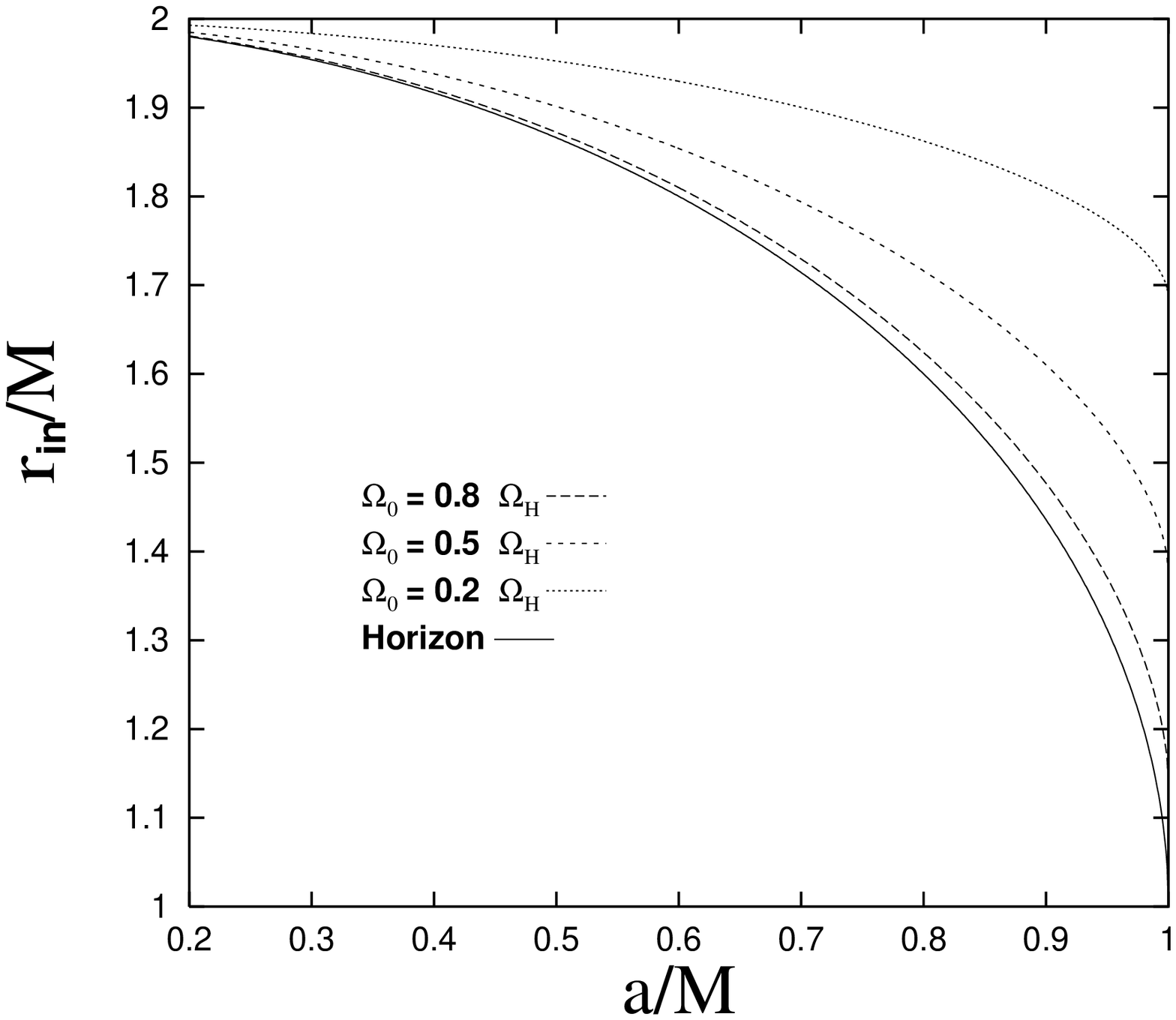, height=8 cm, angle=0}}
\vspace{0.1cm}
Figure 6: The position  of the inner light of surface for the 
Kerr-Newman black hole. $\theta = 0.5 \pi$. 
\end{figure}
\begin{figure}[bh]
\vspace{0.5cm}
\centerline{
\epsfig{figure= 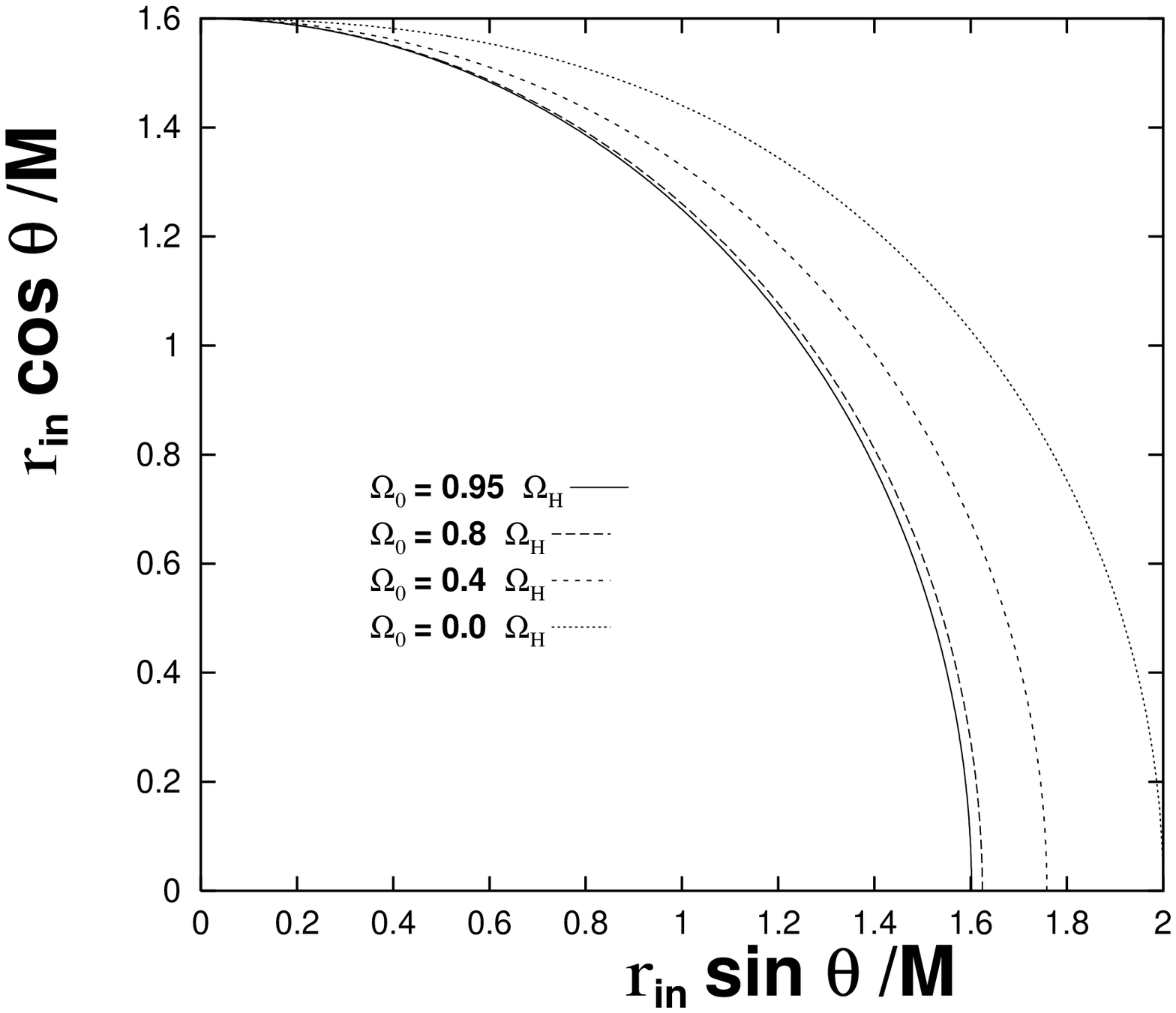, height=8 cm, angle=0}}
\vspace{0.1cm}
Figure 6: The shape  of the inner light of surface for the 
Kerr-Newman black hole.  $a = 0.8 M, e = 0 $. 
\end{figure}

C) {\it the extreme black hole case with} $\Omega_0 =\Omega_H$: 
 For  extreme  Kerr-Newman   black hole case, which occurs when
 $M^2 = a^2 + e^2 $, 
 $g_{tt}^{'}$ at $ \theta = \frac{1}{2} \pi$ 
 is written as 
 \be
g_{tt}^{'}=\frac{M^2}{\Sigma} \frac{y}{1 + y^2}
( x - 1)^2 
\left(x  + 1  - \frac{1}{y} \right)
\left(x  + 1  + \frac{1}{y} \right)
\ee
From this  we  obtain 
 the  position of VLS at $\theta = \frac{\pi}{2} $ as
\bea
  r &=& M ~~~~~~{\rm for ~}   \frac{1}{2} M~\leq a \leq M~ {\rm and }~ a = 0, \\
  r &=& \left( -1 + \frac{M}{a} \right) M  ~~~~~{\rm for ~}  
       0 < a <  \frac{1}{2} M.  
\eea
The second case  corresponds to the extreme  black hole  that
is  slowly rotating and has many charge. (In this case 
 $ e  > \sqrt{3}/2M \approx 0.866 M $). 
In particular in case of $e \leq  \sqrt{3}/2 M $ ( $a = M$  for $ e = 0$)  
the horizon and  the light of velocity surface  
are at the same position. 
 Therefore  in case of  the extreme black hole with $a \geq 1/2 M $
 it is impossible to consider the brick wall model of 't Hooft. 



\section{The Entropy in the Hartle-Hawking Vacuum}

The Hartle-Hawking  vacuum state is one that
the angular velocity $\Omega_0$ is  equal to  that of the  
event horizon, and the temperature $\beta$ is equal to the Hawking
temperature, where the  Hawking temperature and the angular velocity of 
the horizon are  defined  as\cite{Wald}
\be
T_H = \frac{\kappa}{2 \pi},~~~ \Omega_H = \lim_{r \rightarrow r_H}
\left( - \frac{g_{t \phi}}{g_{\phi \phi}} \right).
\ee
Here $\kappa$ is the surface gravity of the  horizon.

First of all let us assume that $\Omega_0 = \Omega_H  $.
In this case, as stated in Sec.4, 
the possible region I is $r_H < r < L < r_{VLS}$.
The outer brick wall  must locate  inside the outer VLS.
This fact was already pointed out by Frolov and Thorne \cite{Thorne}  
to remove the singular structure of the Hartle-Hawking vacuum and
modify it.
Now recall that in general ${g'}_{tt}|_{r = r_H} =0$.
This came from that ${g'}_{tt}$ is the same form as 
$\chi^\mu \chi_\mu = (\xi^\mu + \Omega_H \psi^\mu)(
\xi_\mu + \Omega_H \psi_\mu)$, and $\chi^\mu$ is null on the
horizon.
So it follows that ${g'}_{tt} = ( r - r_H) G(r, \theta)$, where
$G(r,\theta)$ is a non-vanishing  function at  $r = r_H$ except 
the extremal case.
( We can not consider  the extreme black hole case.) 

Therefore for the three black holes the leading behaviors of  the free 
energy $F$  for very small $h$ are then given by 
\bea
\beta F &\approx&  - \frac{N}{\beta^3}
\int d \theta d \phi \int_{r_H + h}^L  dr  \frac{
\sqrt{g_4}}{ ( - g_{tt}^{'} )^2 } \\
&=&  - \frac{N}{\beta^3}
\int d \theta d \phi \int_{r_H + h}^L  dr  \frac{
D(r)}{( r - r_H)^2 G^2(r, \theta) },
\eea
where $D(r,\theta) = \sqrt{g_4}$. Since $D(r,\theta) $ and 
$G(r,\theta)$ are non-vanishing functions  at $r = r_H$ we can expand it
about  $r= r_H$ as follows.
\bea
D(r,\theta) &=& D(r_H,\theta) +  D^{'} (r_H, \theta) (r -r_H) +
O((r- r_H )^2 ),\\
\frac{1}{ G^2(r, \theta)} &=&
\frac{1}{ G^2(r_H, \theta)} + 
\left( \frac{1}{G^2(r_H,\theta)}   \right)^{'} 
+ O((r - r_H)^2 ),
\eea
where $'$ denotes the partial derivative  for  $r$.
So the free energy is approximately given by
\bea
\nonumber
\beta F &\approx & - \frac{N}{\beta^3}
\int d \phi d \theta \int dr 
\left\{     \frac{D(r_H, \theta)}{G^2(r_H,\theta)} 
\frac{ 1}{(r - r_H)^2 } +
\left( \frac{ D(r_H,\theta) }{G^2(r_H,\theta) }   \right)^{'} 
\frac{1}{(r - r_H)}
 + O((r - r_H)^0)  
\right\}   \\
&=& 
- \frac{2 \pi N}{\beta^3} \left\{
\frac{1}{h} \int d \theta 
 \frac{D(r_H, \theta) }{G^2(r_H,\theta) } 
- \ln (h) \int d \theta 
\left( \frac{ D(r_H,\theta) }{G^2(r_H,\theta) }   \right)^{'} 
+  ...    \right\},
\label{gen}
\eea
which show that 
generally, in addition to the linear divergence term in $h$, 
there is a logarithmic one in the case of  rotating black hole.
If we  written the free energy  in terms of the proper distance 
cut-off $\epsilon$, it become  very  simple  form.
 \bea
 \nonumber
\beta F  &\approx& -  
\frac{N}{ \beta^3} \int_{r = r_H} d \phi d \theta 
\sqrt{g_{\theta \theta} g_{\phi \phi}} 
 \int_{r_H + h}^L dr \sqrt{g_{rr}}
\left( \frac{g_{\phi \phi}}{g^2_{t \phi} -g_{tt}g_{\phi \phi}}
\right)^{3/2}   \\
&\approx& -  \frac{N}{ 2 (  \kappa \beta)^3 } 
      \frac{A_H}{\epsilon^2},
              \label{free2}
\eea
where $A_H$ is the area of the event horizon,
 and $\epsilon$ is the 
proper distance from the horizon to $r_H + h$.
\be
\epsilon = \int_{r_H}^{r_H + h} dr \sqrt{g_{rr}}.
\ee
However the proper distance cut-off is dependent on the 
coordinate $\theta$, which is the general property of the 
rotating black hole.

From the free energy $F$  we obtain the leading behaviors of 
the entropy $S$  as
\bea
\nonumber
S &=& \beta^2 \frac{\partial}{\partial \beta} F   \\
 &\approx&    
 \frac{N}{\beta^3} \left(  A ~ \frac{1}{h}   + B \ln (h)  + finite 
 \right),
\eea
where $A$ and $B$ are in $c$-number in Eq.(\ref{gen}),
 or
\be 
S \approx    \frac{4 N}{ 2  ( \kappa \beta)^3 } 
\frac{A_H}{\epsilon^2}. 
\label{Entropy} 
\ee
The entropy $S$  is  linearly and logarithmically divergent 
as $h \rightarrow 0$. 
The divergences arise because   the density of state for a given $E$
 diverges as $h$ goes to zero.

Now  we take $T$ as the Hartle-Hawking temperature 
$T_H = \frac{ \kappa}{2 \pi}$. 
Then the  entropy  becomes
\be
S_H  \approx  \frac{N 8 \pi^3 }{\kappa^3} \left(
A~ \frac{1}{h}  +  B \ln (h) + finite \right),
\ee
or
\be
S_H \approx  \frac{N}{4 \pi^3}
\frac{ A_H}{ \epsilon^2}.   \label{result}  
\ee
The entropy of a scalar field  in Hartle-Hawking state 
diverges quadratically  
in $\epsilon^{-1}$ as the system approaches  the horizon.
Or it diverges in $h^{-1}$ and $ \ln (h) $.
In case $ a= 0$ our result (\ref{result}) agrees with the   
result calculated by 't Hooft \cite{tHooft} and 
with one in ref.\cite{ohta}.
These facts imply that the leading behaviors of entropy  (\ref{result})
is general form.


\section{Summary and Conclusion}

By using the brick wall method we have calculated the entropies  
of the rotating systems with a rotation $\Omega_0$ 
at thermal equilibrium with temperature $T$ in the  
rotating  black holes.  
In WKB approximation to get the real finite free energy and entropy 
the system must be in the region I.
As the system approaches the VLS ( $r_{in}$ and $r_{VLS}$)
the thermodynamic quantities become divergent. From this fact 
{\it we conclude that the divergence of the thermodynamic quantities 
including the entropy is related to the stationary limit
surface in the  co-moving frame}. In spherical symmetric black hole
the stationary limit surface and the  event horizon are coincide.
Only when $\Omega_0 =\Omega_H$ the system can be approach the horizon.
The entropy for this case is linearly and logarithmically divergent 
as the ultraviolet cut-off goes to zero.
To remove  such a divergence, in addition to the renormalization of the 
gravitational constant,  we need the renormalization of the 
curvature square term\cite{ohta}. But after the renormalization
the entropy  does not proportional to the area of the event  horizon.
If we use the proper distance cut-off 
the entropy is proportional to the  horizon area $A_H$. But 
the cut-off  depends on the  coordinate $\theta$.

Another particular point is that in the extremal black hole case
we can not consider the brick wall method of 't Hooft except for
the case $ 0 < a < 1/2 M$ in Kerr-Newman black hole.

\section*{Appendix} 
For the three rotating  black holes  the metrics, the surface 
gravities, , and the proper distances $\epsilon$ 
are given as follows:

1)  the Kaluza-Klein  black hole \cite{Frolov} 
\begin{eqnarray}
\nonumber
ds^2 &=& - \frac{\Delta - a^2 \sin^2 \theta }{B \Sigma}dt^2
     -2 a \sin^2 \theta \frac{1}{\sqrt{1 - v^2 }} 
     \frac{Z}{B} dtd \phi  \\
     & & ~ + \left[
     B \left( r^2 + a^2 \right)  + a^2 \sin^2 \theta 
     \frac{Z}{B} \right]
     \sin^2 \theta d \phi^2 + 
     \frac{ B \Sigma}{\Delta} dr^2 + B \Sigma d \theta^2,
\end{eqnarray}
where
\begin{equation}
\Delta = r^2 - 2 \mu r + a^2 ,~~~
\Sigma = r^2 + a^2 \cos^2 \theta,~~~
Z = \frac{2 \mu r}{\Sigma},~~~
B = \left( 1 + \frac{v^2 Z}{1 - v^2 } \right)^{\frac{1}{2}}.
\end{equation}
The physical mass $M$, the charge $Q$, the angular momentum $J$,
and the horizon
are expressed by the parameters $v,\mu,$ and $a$ as
\be
M = \mu \left[
     1 + \frac{ v^2 }{2 ( 1 - v^2 )} \right],~~~
     Q = \frac{\mu v}{ 1 - v^2 },~~~
     J = \frac{\mu a}{\sqrt{1 -v^2}},
     r_H = \mu + \sqrt{ \mu^2 - a^2}.
\ee
The surface gravity and proper distance are
\begin{eqnarray}
\kappa_{Kaluza-Klein} &=& 
       \frac{ \sqrt{( 1 - v^2) ( \mu^2 - a^2)}}{ r_H^2 + a^2},\\
\epsilon_{Kaluza-Klein} &=& 
    2  \left( \frac{B(r_H) \Sigma (r_H) }{ 2 r_H - 2 \mu} \right)^{1/2}
      \sqrt{h}.    
\end{eqnarray}

2) the Sen black hole\cite{sen}:
\bea
ds^2 &=&  - \frac{\Delta - a^2 \sin^2 \theta}{\Sigma}dt^2
- \frac{4 \mu r a \cosh^2 \gamma \sin^2 \theta}{\Sigma} dt d \phi \\
& &~+ \frac{\Sigma}{\Delta} dr^2 + \Sigma d\theta^2
+ \frac{\Lambda}{\Sigma} \sin^2 \theta d \phi^2,
\eea
where
\bea
\Delta &=& r^2 - 2 \mu r + a^2 ,~~~
\Sigma = r^2 + a^2 \cos^2 \theta + 2 \mu r \sinh^2 \gamma, \\
\Lambda &= & \left( r^2 + a^2 \right)  \left(
  r^2 + a^2 \cos^2 \theta \right) + 2 \mu r a^2
  \sin^2 \theta  \\
  & &~  + 4 \mu r \left( r^2 + a^2 \right) \sinh^2 \gamma 
   + 4 \mu^2 r^2 \sinh^4 \gamma.
\eea
The mass $M$, the charge $Q$, the angular momentum $J$,  and the
horizon are given 
by parameters $\mu,\beta$, and $a$ as
\be
M = \frac{\mu}{2} \left( 1 + \cosh 2 \gamma \right),~~
Q = \frac{\mu}{\sqrt{2}} \sinh 2 \gamma,~~
j = \frac{ a \mu }{2} \left( 1 + \cosh 2 \gamma \right),~~
r_H = \mu + \sqrt{\mu^2 - a^2}.
\ee
The surface gravity and proper distance are
\begin{eqnarray}
\kappa_{Sen} &=& 
          \frac{ \sqrt{ ( 2 M^2 - e^2 )^2 - 4 J^2 }
		    }{ 2 M \left[
	 2 M^2 - e^2 + \sqrt{ ( 2 M^2 - e^2 )^2 - 4 J^2 }
			   \right]  }, \\
\epsilon_{Sen} &=& 
 2 \left( \frac{ r_H^2 + a^2 \cos^2 \theta + 2 \mu r_H \sinh^2 
           \gamma }{ 2 r_H - 2 \mu } \right)^{1/2}  \sqrt{h}.
\end{eqnarray}

3)  the charged Kerr black hole \cite{kerr}
\begin{eqnarray}
\nonumber
ds^2 &= & - \left( 
		\frac{ \Delta - a^2 \sin^2 \theta }{\Sigma}
		\right) 
		dt^2 - \frac{2 a \sin^2 \theta ~( r^2 + a^2 - \Delta)}{
		\Sigma } dt d\phi \\
    & &~ + \left[ \frac{(r^2 +a^2 )^2 - \Delta a^2 \sin^2 \theta }{
  \Sigma}  \right] \sin^2 \theta d \phi^2 + \frac{\Sigma}{\Delta} dr^2 +
    \Sigma d \theta^2, 
\end{eqnarray} 
where 
\begin{equation}
\Sigma = r^2 + a^2 \cos^2 \theta, ~~~~~  
\Delta = r^2 + a^2 + e^2 - 2 M r,
\end{equation}
and $e,a,$ and $M$ are charge, angular momentum per unit mass, and
mass of the spacetime respectively. 
The event   horizon  is 
\be
 r_H =  M + \sqrt{M^2 - a^2 - e^2 }.
\ee
The surface gravity and proper distance are
\begin{eqnarray}
\kappa_{Kerr}&=& \frac{ \sqrt{M^2 - a^2 -e^2}}{ 2 M 
	\left[ M + \sqrt{M^2 - a^2 - e^2} \right] - e^2},   \\
\epsilon_{Kerr} &= &
 2 \left( \frac{ r_H^2 + a^2 \cos^2 \theta }{ 2 r_H - 2 M } 
\right)^{1/2} \sqrt{h}.   
\end{eqnarray}

\begin{flushleft}
{\bf Acknowledgment}
\end{flushleft}

This work is partially supported by Korea Science and Engineering 
Foundation.


\end{document}